\documentclass{sigchi}

\usepackage{balance}  
\usepackage{graphics} 
\usepackage{times}    
\usepackage{url}      

\makeatletter
\def\url@leostyle{%
  \@ifundefined{selectfont}{\def\UrlFont{\sf}}{\def\UrlFont{\small\bf\ttfamily}}}
\makeatother
\def\pprw{8.5in}
\def\pprh{11in}

\usepackage[pdftex]{hyperref}
\hypersetup{
pdftitle={SIGCHI Conference Proceedings Format},
pdfauthor={LaTeX},
pdfkeywords={SIGCHI, proceedings, archival format},
bookmarksnumbered,
pdfstartview={FitH},
colorlinks,
citecolor=black,
filecolor=black,
linkcolor=black,
urlcolor=black,
breaklinks=true,
}

\begin{document}

\title{Botivist:\\Calling Volunteers to Action using Online Bots}

\numberofauthors{3}
\author{
  \alignauthor Saiph Savage\\
    \affaddr{West Virginia University \& }\\
    \affaddr{National Autonomous University of Mexico}\\
    \email{norma.savage@mail.wvu.edu}\\
  \alignauthor Andres Monroy-Hernandez\\
    \affaddr{Microsoft Research}\\
    \email{amh@microsoft.com}\\
  \alignauthor Tobias Hollerer\\
    \affaddr{UC Santa Barbara}\\
    \email{holl@cs.ucsb.edu}\\
}

\maketitle

\begin{abstract}
To help activists call new volunteers to action, 
we present Botivist: a platform that uses Twitter bots to find potential volunteers and request contributions. By leveraging different Twitter accounts, Botivist employs different strategies to encourage participation. We explore how people respond to bots calling them to action using a test case about corruption in Latin America. Our results show that the majority of volunteers ($>80\%$) who responded to Botivist's calls to action contributed relevant proposals to address the assigned social problem. The number of contributions and their relevance varied by the strategy used. Some strategies that work well offline and face-to-face appeared to hinder people's participation when used by an online bot. We analyze user behavior in response to being approached by bots with an activist purpose. We also  provide strong evidence for the value of this type of civic media, and derive design implications.   
\end{abstract}

%


\section{Introduction}
Activist groups usually have a small set of highly motivated core members who give a great deal of their own time and resources to make change in the world~\cite{massung2013using} -- for example, to fight corruption.
However, to achieve their goals, activists cannot rely entirely on these core members. They usually require a larger crowd of volunteers who believe in the cause and can perform small actions, e.g. individuals around the world who can help report corruption in their local area~\cite{Starbird:2011:VSD:1978942.1979102}. Activist groups need the support of casual volunteers: a group of interested, but less committed individuals~\cite{smith1981altruism}.

For years, activists went door-to-door to recruit and engage casual volunteers. Recently, new technologies have helped activists build the support of casual volunteers~\cite{persaMersa,massung2013using}. Some use mailing lists to maintain continuous communication with their volunteers~\cite{saeed2011analyzing,sommerfeldt2011activist}. Others are using social media~\cite{velasquez2014youth}. Facebook has been particularly useful to recruit and coordinate volunteers, issuing calls to action for fundraisers or demonstrations~\cite{voida2012bridging,wulf2013fighting}. Twitter has also enabled activist groups to raise awareness and mobilize even those unaffected by the activists' cause or who live in distant regions~\cite{Starbird:2011:VSD:1978942.1979102,starbird2011voluntweeters}. 

However, despite the technological advancements, social computing has not been widely employed to connect casual volunteers~\cite{voida2012bridging}. Many eager individuals receive little direction on how to help~\cite{Starbird:2011:VSD:1978942.1979102}. Current technologies also do not help activists mobilize people. Activist groups must still spend time figuring out how they will present their campaigns in order to successfully trigger action~\cite{velasquez2014youth}.
\begin{figure}
  \begin{center}
  \includegraphics[width=200pt]{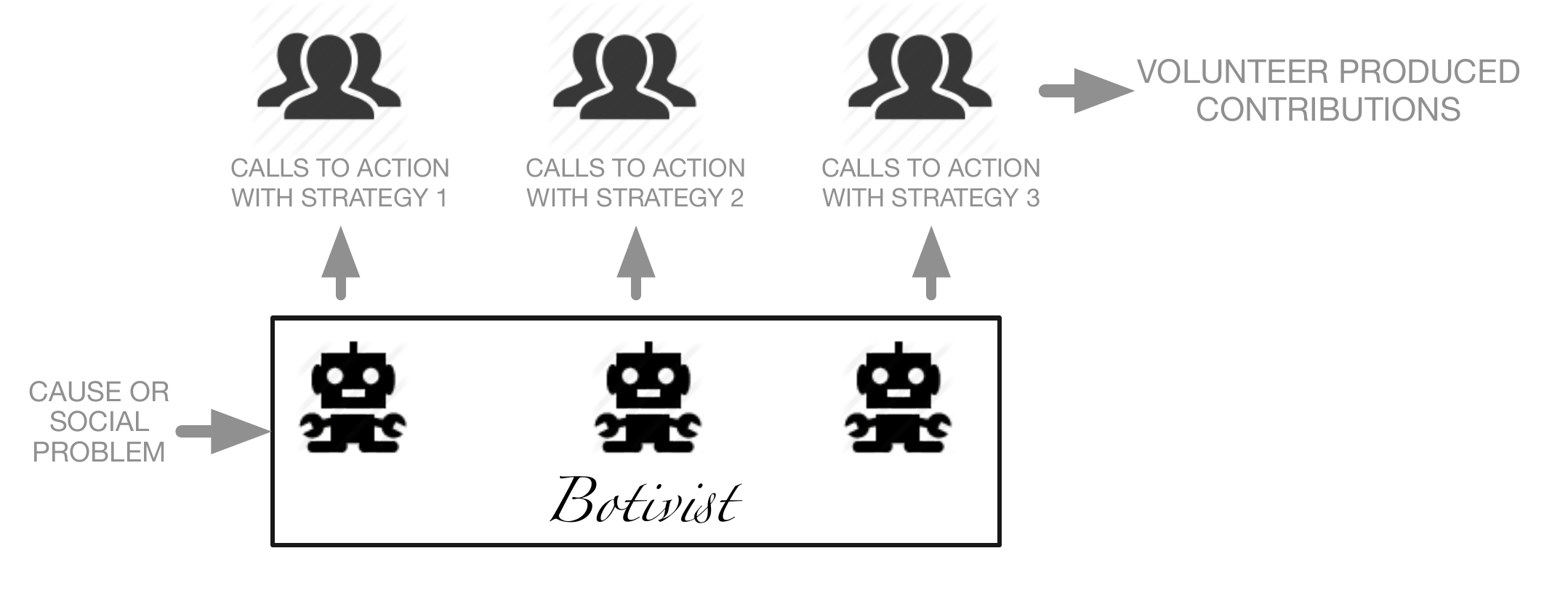}
  \caption{ Overview of Botivist: Activists first provide the social problem for which they want action. Botivist then tries different strategies to  trigger contributions from volunteers.}
  \label{fig:botSystem}
  \end{center}  
\end{figure}

Understanding how the presentation of a campaign (message) affects the engagement of volunteers and people in general has been extensively studied in both theories for civic engagement and marketing~\cite{fogg2002persuasive,levin1988information,maheswaran1990influence,rettie2005text,tversky1986rational}. There has also been a growing interest in exploring how technology can be used to frame messages better and more persuasively~\cite{fogg2002persuasive}. Companies, organizations, and even governments have taken to social media to reach a wider audience and influence behavior~\cite{persaMersa}. Some have gone as far as to set up fake accounts operated by automated software (bots) in order to feign support and influence real users~\cite{forelle2015political}. In Mexico's 2012 presidential election, the winning party used over 10,000 bots to swamp online discussion~\cite{mexElec}.

Large and well-funded organizations have devised complex social media strategies to get their message across. However, most activist groups are still in the dark, trying to find the best strategies to mobilize people~\cite{voida2012bridging}. The main difficulty comes from the fact that volunteers' level of agency changes based on how they perceive the efforts of activists~\cite{velasquez2014youth}. But, social media has transformed how people interpret activists' varying messages and consequently their efforts~\cite{saeed2011analyzing}. It can thus be very difficult for activist groups to keep up with new technology and to predict the outcomes of adopting a certain strategy. This complexity has forced many activists to limit who among their organization can use social media~\cite{voida2012bridging}. They usually prefer to have a ``point person'' in charge of the group's social media strategy. However, even when the point person finally ``figures out'' the best ways to trigger participation, it is not easy to transfer that knowledge to others~\cite{saeed2011analyzing}. Additionally, it can be costly or impossible for the person to present the group's campaigns differently to test what is the most effective. All of this hampers activists' success.

To help activists identify the best strategies, we present Botivist, a platform that, by leveraging online bots, allows activist groups to try different strategies for calling volunteers to action. Figure \ref{fig:botSystem} presents an overview of Botivist. The group first presents the cause for which they want to generate action. The platform then tries different strategies to request contributions on social media, and help advance the group's plans. Activist groups thus receive help about how to prompt contributions from strangers, reducing the need to invest time in people who might never contribute.

The purpose of this paper is twofold. First, we aim to analyze the opportunities, limitations, and challenges of a system that uses online bots to call people to action in order to act upon social issues. Secondly, we analyze the online behaviours of those who are most likely to respond to activist bots. More specifically, (1) we focus on understanding how social media users respond to differently-framed messages (strategies) when delivered by an activist bot. 
Most previous work studied messages framed primarily when the information came from a company, organization, or individual, and has not addressed how people respond when the messages are presented by an online bot, especially for the topic of activism. (2) In addition, an effort is made to understand the communication patterns of the people who decide to engage with Botivist. This analysis is important as these individuals could one day become core activists.

For this purpose, we designed and conducted experiments on Twitter. We designed Botivist as a Spanish speaking agent because we had more background with the language and its usage  in activism: Botivist's main researchers are native Spanish speakers, with knowledge and experience on activism in Latin America; we also had direct access to activists from Mexico\footnote{We had continuous discussions with activists from Mexico's Revolutionary Action Group (GAR) who collaborate with activists' across the world to fight corruption and bring justice to workers ~\url{https://es.wikipedia.org/wiki/Grupo\_de\_Accion\_Revolucionaria}} who helped polish Botivist's messages. We focused our calls to action on corruption and corrupt officials' impunity from prosecution. These are some of the most pressing social issues fought in general by activists in Spanish speaking countries, especially those in Latin America~\cite{pewLatinAmerica,impunityLatinAmerica}. 

Our aim was for Botivist's calls to action to facilitate collaboration. We chose this focus because activist groups usually need volunteers to work together towards a common goal. Thus, Botivist's tweets were structured to create collaboration by mentioning three users and suggesting that they collaborate to fight a social problem.  Mentioned users might not know each other, but they would all be currently using similar keywords, e.g., in their latest tweets all of them mentioned ``corruption''. By suggesting collaborations, Botivist can also help people to meet and connect.

To assess and compare the reactions that Botivist triggers in volunteers, we use methods from previous literature on reactions to different strategic presentations of a message~\cite{levin1988information,maheswaran1990influence,rettie2005text,tversky1986rational}, as well as research that analyzed the quality of information from strangers on social media~\cite{nichols2013analyzing}. We deployed our platform publicly on Twitter, where our activist-designed and controlled bots invited people to organize around particular social issues using different strategies. 175 volunteers responded to Botivist's calls to action. These volunteers made 1,236 contributions (424 tweets, and 813 favorites and retweets). We found that Botivist's most effective strategy for prompting contributions from volunteers was to be upfront and direct. Interestingly, when Botivist used techniques designed to be persuasive and known to be effective in direct human-to-human interaction, it received far fewer replies. At times these techniques even seemed to prompt individuals to discuss whether bots should help solve social problems. We found that by being less openly persuasive in its message, Botivist encouraged almost double the number of replies. Our results also indicate that those who decided to engage with Botivist were individuals who consistently tweeted terms related to activism and politics, with a small sub-population of users who tweeted terms related to marketing analytics. 

Together, our results highlight the importance of understanding a community before simply adopting persuasive technology and expecting it to function. However, the benefit of Botivist is that by allowing activists to systematically test different strategies, it might not be necessary to understand a community in detail. Our work shows the strength of this new type of civic media.

\section{Related Work}
\subsection{Botivist: Strategically Framed Messages}
There is extensive literature on how a message's framing influences people's preferences for a product, or how much they participate in an event~\cite{levin1988information,maheswaran1990influence}. Such  research has uncovered the power of presenting something in terms of its positive benefits or negative drawbacks can have. Framing can influence listeners' opinions and levels of participation, making people more likely to buy a product if its ad highlights relevant benefits, than if the ad lacks such information. Related work has also identified the value of integrating particular messages into a call to action to motivate participation.  Shen~\cite{shen2010mitigating} points out that integrating a solidarity component into the requests made to volunteers led to volunteers completing their tasks more thoroughly. 

``Persuasive computing''~\cite{fogg2002persuasive} has combined this literature with  technology to demonstrate how computation can be used to influence people's behavior~\cite{persaMersa}. Many companies and organizations are now also using social media technology to influence choices and shape the habits of their consumers~\cite{rettie2005text}, even their health choices~\cite{Baumer:2012:PPO:2145204.2145279}. Online platforms have also adopted persuasive computing to motivate greater participation. Ling et al.~\cite{ling2005using} found that by simply reminding people of their connection to an online community, people posted more. However, most persuasive computing research has paid little attention to designing technology to enhance activism. 
While there has been growing interest in creating systems to help activists coordinate volunteers~\cite{massung2013using,brady2015gauging} most of these systems are not designed to audience-test the framing of different messages. By having different strategies for calling people to action, individuals can be persuaded to participate more in a social cause. Botivist gives activists and organizers flexibility in framing the most effective call to action.



\subsection{Botivist: Online Automated Agents}
Since the 1960s, there has been a proliferation of automated conversational agents~\cite{Weizenbaum:1966:ECP:365153.365168}, and it is now commonplace for humans to respond to them~\cite{nass1994computers}. We designed Botivist to behave in line with other mainstream conversational agents. Most previous work on online bots has focused more on developing techniques to detect these agents~\cite{forelle2015political}. However, this ``Turing test'' approach has left out the possibility that bots can persuade and mobilize people on social media. This potential has not yet been fully understood, let alone explored in the field of activism. Aiello et al.~\cite{aiello2012people} began an exploration in the area by setting up a social experiment that examined how people interact with a bot in an online community. The work uncovered how a bot could acquire social relevance even when using simple canned responses and lacking profile information. Botivist builds on these findings to create online social agents that start to mobilize people to help activist groups. 
\section{Botivist}
Botivist is a web application that activist groups can use to call volunteers to action to help address a social issue. 
We based Botivist on the well-known idea that by simply changing how a message is presented, it is possible to trigger diverse reactions~\cite{levin1988information,maheswaran1990influence,rettie2005text,tversky1986rational}. Therefore, by simply framing a call to action differently, we can prompt different amounts and kinds of responses. However, deciding how to frame a call to action to obtain a desired turnout is not simple. Therefore Botivist probes different strategies, i.e., ways of presenting a call to action. 
\subsection{Botivist Strategies}
We selected some of the most common strategies used by activists to call volunteers to action~\cite{fleishman1980collective}. The  strategies are also known to be some of the most effective to garner participation~\cite{shen2010mitigating,tversky1986rational}. The messages used for each strategy were written by the first author, and polished with the help of real activists. The activists discussed the messages, and then corrected the language to produce content that they would use. 

To try out each strategy, Botivist has a set of Twitter  accounts to communicate with potential volunteers.  All of the accounts identify themselves exactly the same. They mention they are bots (with a bio that states they are a bot; their profile picture is also of a bot.) All accounts also have at least 50 followers. The only main difference across accounts is the strategy used to communicate and call volunteers to action. 
\subsubsection{Direct Strategy} This strategy is about being upfront and direct when making petitions~\cite{Aleahmad:2008:FSE:1358628.1358801}. When using this strategy, Botivist directly calls people to action to find solutions for a social issue. For example:~\emph{``Could we collaborate to brainstorm solutions to the problem of corruption?''}

\subsubsection{Solidarity Strategy} This strategy posits that feelings of solidarity or empathy can drive people to respond to a call to action~\cite{fleishman1980collective,shen2010mitigating}. When using this strategy, Botivist directly calls people to action, but it also shares solidarity quotes. For instance, a sample call to action is:~\emph{``Could we collaborate to brainstorm solutions to the problem of corruption?'' ``Remember, that: One for all, all for one!''} Note that in this strategy for each call to action or reply, two tweets are sent: one tweet soliciting a reply, the same as the direct strategy, and one tweet to share the solidarity quote.

\subsubsection{Gain Strategy} This strategy is about presenting a calls to action in terms of the gain that people would receive if they participate~\cite{tversky1986rational}. These calls to action also mimic those used in the direct strategy, but add additional text to emphasize the gain. A sample call to action from this strategy:~\emph{``Could we collaborate to brainstorm solutions to the problem of corruption? We might improve our cities!''}

\subsubsection{Loss Strategy} This strategy is about presenting a call to action in terms of the loss that people would receive if they do not participate~\cite{tversky1986rational}. A sample call to action from this strategy:~\emph{``Could we collaborate to brainstorm solutions to the problem of corruption? If not, our cities might suffer!''}

We envisioned activist groups using Botivist mostly at the beginning of a collective project. Botivist thus focuses on prompting action in the form of discussions or plan proposals to help solve the social problem the activists are focused on, just as discussion and brainstorming are among the first steps of any collective effort.  All of Botivists' responses echo  the ``How Might We'' questions used to launch brainstorming~\cite{carroll2010destination}. Depending on which strategy Botivist is pursuing, it will have a slightly different follow-up question. For instance, when following the gain strategy, Botivist asks for ideas, and then states that by participating in brainstorming, people could help improve their cities; whereas when following the solidarity strategy, Botivist will ask for ideas, and share a solidarity quote at the end. Table \ref{table:questionsBot} presents the list of follow-up questions for the direct strategy. The follow-up questions of the other strategies are exactly the same, except for the added text. For instance, the gain strategy had the question:~\emph{How do we fight corruption in our cities \& thus improve them?} All calls to action and followup messages for brainstorming were verified and polished by real activists. A strategy has one particular call to action associated and a list of replies to pursue brainstorming. Botivist randomizes its replies. Botivist only targets a person once, and engages further only if it receives a response.  

\begin{figure}
 \begin{center}
  \includegraphics[width=180pt]{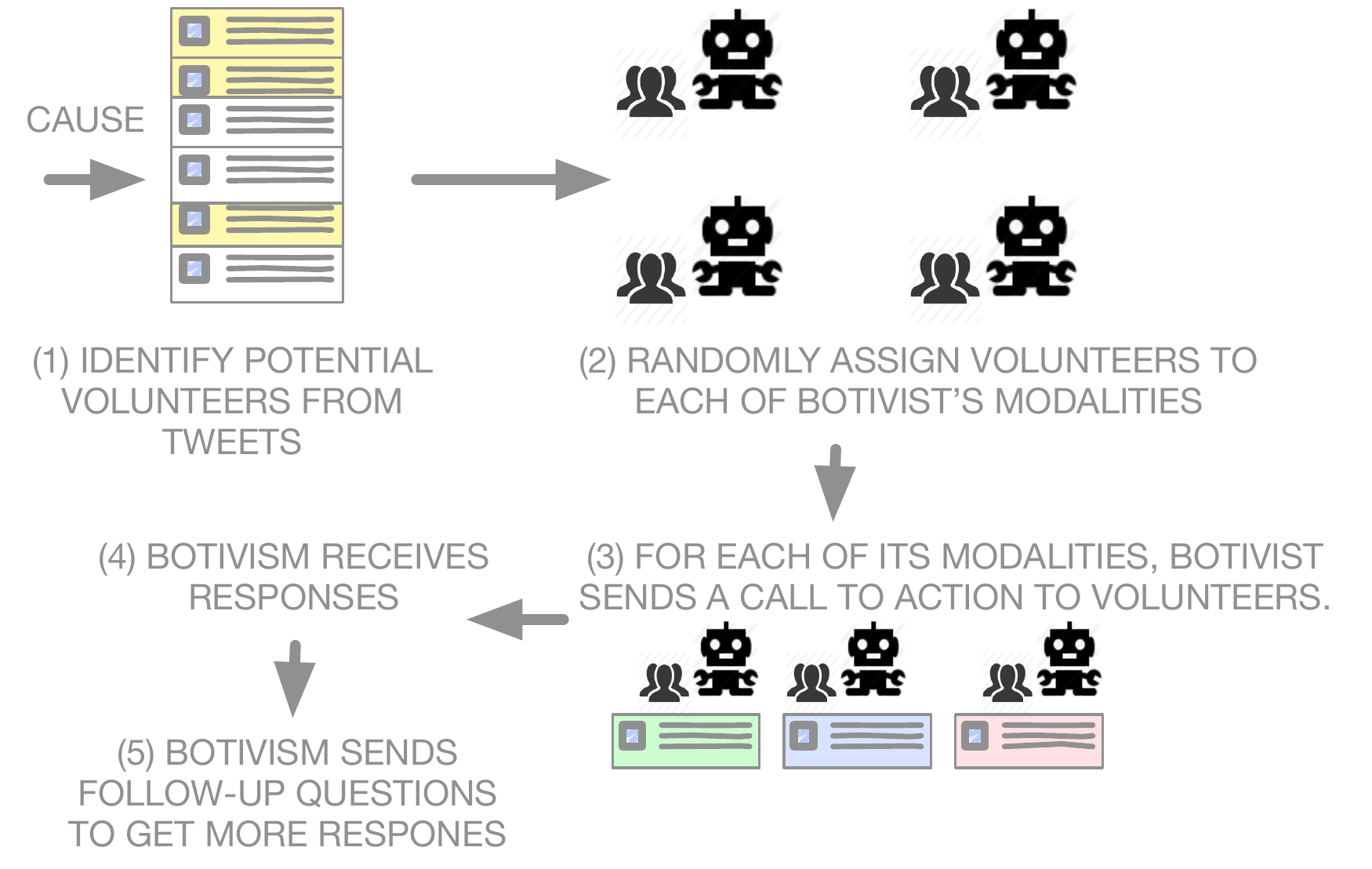}
  \caption{Botivist's process for calling volunteers to action and obtaining responses.}
  \label{fig:botSYS}
  \end{center}  
\end{figure}
\subsection{Botivst Operation}
An activist group first defines the social problem for which they seek volunteers -- in our test case, ``corruption.'' Botivist identifies a set  of potential volunteers and targets them by mentioning them in a tweet.  Botivist then waits for replies, in which case it uses the opportunity to request more responses. We set Botivist up to target the same number of volunteers for each strategy category. Figure~\ref{fig:botSYS} shows Botivist's work flow. 

To determine who will be mentioned in a tweet, we take into account the findings from previous work on social media interactions between strangers~\cite{nichols2013analyzing} and research on publicly mentioning people in social media content~\cite{Savage:2015:TMM:2700171.2791055}. Nichols et al.~\cite{nichols2013analyzing} showed that obtaining responses to questions from  strangers usually had a low response rate,  even under favorable conditions. We thus designed Botivist to maximize the chances that people would respond to the call to action. Botivist targets individuals who have just publicly tweeted something related to the activists' social problem of interest, based on simple keyword matching. For instance, if the activist group wants volunteers to fight corruption, the system will target people who have just publicly shared a tweet explicitly mentioning corruption. Group managers can provide a list of relevant keywords or hashtags to the system. This method is drawn from how activists operate online, searching for relevant keywords and making publicly addressed tweets to seek a wide range of help and build new social connections~\cite{Starbird:2011:VSD:1978942.1979102}. Similar to real activists, Botivist engages people across countries, and has no prejudice against people based on their initial support or opposition to the social issue at hand~\cite{starbird2012will}. 

Once a relevant tweet is detected, its author is randomly assigned to one of Botivist's strategies. When a strategy has three authors (persons) assigned, Botivist sends a strategic tweet that mentions and calls the individuals to action to work together on the issue. Note that Botivist mentions multiple users in a tweet to facilitate cooperation among possible volunteers. Given Twitter's space constraints, three users was the maximum that Botivist could mention in one tweet. Botivist is also constantly monitoring Twitter for responses to its calls to action. Botivist replies with canned responses that depend on the strategy Botivist used to communicate with the group. The follow-up messages all focus on prompting action from the volunteers.

\begin{table}[htdp]
\begin{tabular}{l*{6}{c}}
\small Q\# & \small {\bf Question Text}\\
\hline
\small & \small \emph{Direct Strategy}\\ 
\hline
\small 1 & \small How do we fight corruption in our cities?\\
\small 2& \small How do we fight corruption in our countries?\\
\small 3 & \small How do we use Twitter to fight corruption?\\
\small 4&  \small How do we use the people to fight corruption?\\
\small 5&  \small What should we change personally to fight corruption?\\
\small 6&  \small What should we reduce to fight corruption?\\
\small 7&  \small  What should we change at home to fight corruption?\\
\hline
\end{tabular}
\caption{Questions asked in the direct strategy.}
\label{table:questionsBot}
\end{table}


\section{Evaluation}
This paper hypothesizes that online conversational bots can be used to call volunteers to action to work on solving a social issue. We focus our evaluation on the two main aspects of this claim. First, for a given social problem, is it possible to use bots to call volunteers to action? Can Botivist obtain responses from strangers on social media, and what solicitation strategies are most conducive for obtaining results?  Second, can Botivist prompt these volunteers to produce relevant contributions? Does Botivist trigger volunteers to discuss the issue or to propose plans? Similar to \cite{nichols2013analyzing}, our goal is to shed light on the type of contributions which autonomous agents can assemble in short bursts of times to get more immediate usable results.

To answer these questions, we used Botivist to publicly call volunteers to action regarding the social problems of corruption and impunity~\cite{pewLatinAmerica,impunityLatinAmerica}. 

We were able to attract 175 volunteers over the course of two days. Because we did not group targeted users based on their location, Botivist likely called to action people from different countries, similarly to how real activists' Twitter campaigns function~\cite{starbird2012will}. 

\subsection{Method: Botivist and Volunteers' Participation}
Botivist was active from April 26th until May 7th, 2015. We used the keywords ``corrupcion" and ``impunidad" (Spanish for corruption and impunity) to select users with relevant tweets. For both social problems, we loaded calls to action and follow-up questions specific to each of Botivist's strategies.  Botivist alternated between social problems, first calling a group to fight corruption, and subsequently calling another group to fight impunity. To minimize any sequence effect, Botivist tried to make its calls to action for all strategy types almost at the same time. However, we randomized the delay time between Botivist's calls to action to avoid being labeled as spam by Twitter. The researchers also constantly monitored Botivist's Twitter accounts to ensure they were still running and not blacklisted. 
All tweets received and sent by our system were collected in a database for further  analysis.

\subsection{Results}
\begin{table}[htdp]
\small{
\begin{tabular}{l p{1cm} c c c c}
  & Total & Direct & Loss & Gain & Solidarity \\
\hline
Calls to Action & 376 & 94 & 94 & 94 & 94 \\
Followup Questions & 557 &  158 &  80 & 79 &  120\\
Volunteers &  175 &  94 &  31 &  27 &  23\\
Volunteer Replies &  423 &  204 &  53 &  74 &  92\\
Reply Rate & 45 \% &  81\%& 30\% & 43\% & 21\%\\
Interactions Bot & 320 &  90& 48 & 57 & 250\\
Interactions Volunteers  & 493 &  274& 71 & 85 & 62\\
\hline
\end{tabular}
}
\caption{ Summary of Botivist's and volunteers' replies and interactions. Total column shows the results summed across strategies}
\label{table:valores}
\end{table}

\begin{figure}
\begin{center}
\includegraphics[width=180pt]{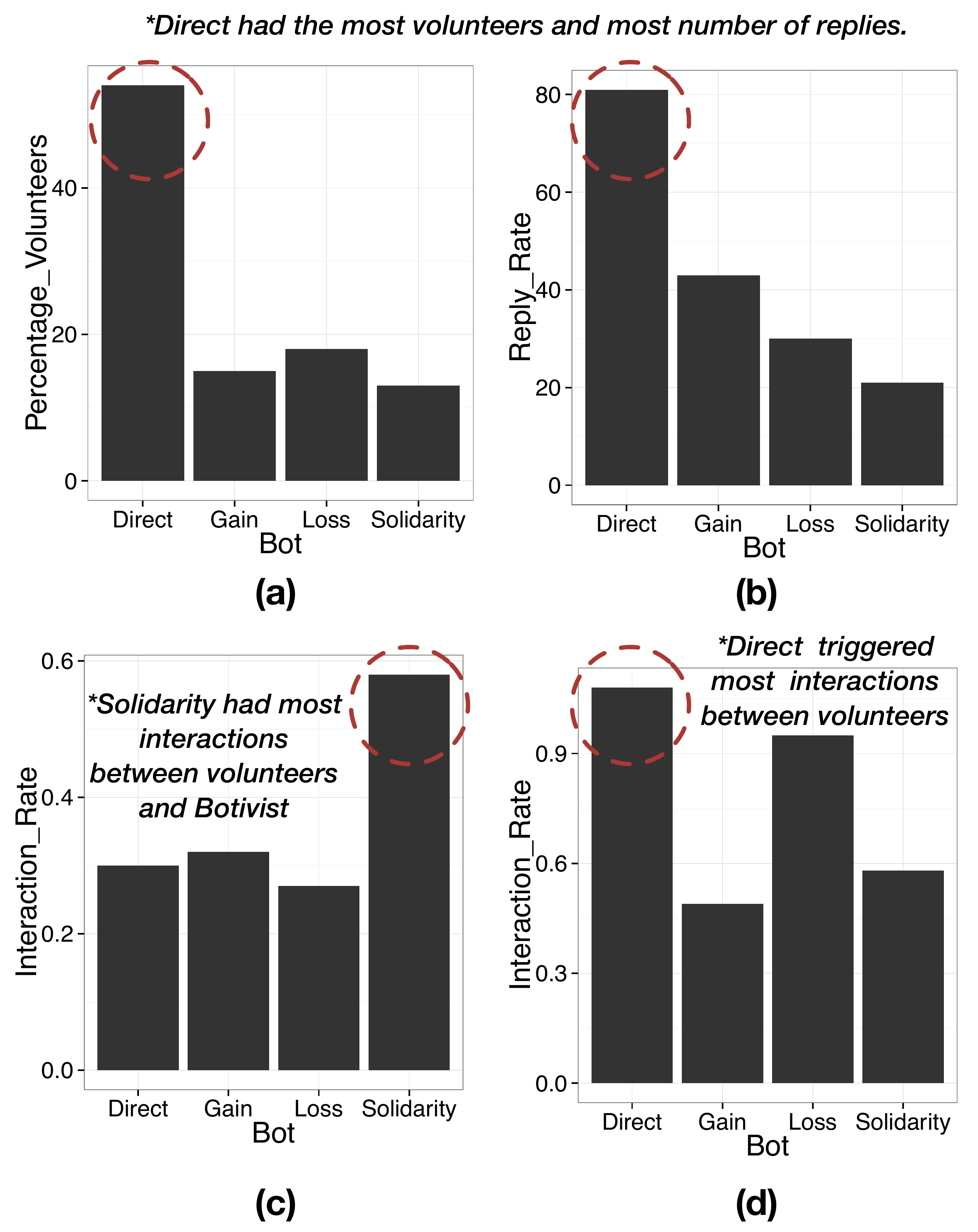}
  \caption{ Overview of the number of volunteers and responses that each strategy triggered.  The direct strategy had the highest participation and prompted the most responses.    
  }
  \label{fig:volunteers}
  \end{center}  
\end{figure}
Botivist called to action 376 groups of three: 94 groups per strategy (47 for corruption; 47 for impunity). In total, Botivist received replies from 175 volunteers. These volunteers made 1,236 responses or other actions (424 tweets; 320 retweets and favorites to Botivist's content; 493 retweets and favorites to the content of other volunteers.) Table \ref{table:valores} shows a breakdown of the replies and interactions of Botivist and volunteers. Figure~\ref{fig:volunteers} shows an overview of how people responded to Botivist's calls to action. The number of participants varied across strategies, as well as their contributions.

Figure \ref{fig:volunteers}a shows the percentage of people who responded to each strategy. There was a significant difference in the number of unique contributors across strategies, based on an ANOVA [$F(3,174)=38.94$, $p<0.001$].
The direct strategy was the most effective at mobilizing unique contributors. Over 50\% of Botivists' participants were gathered via this strategy. By being direct and upfront, Botivist was able to obtain responses from over 30\% of the people it targeted. This compares favorably with research that shows only 21\% of social media users have participated in groups involved in political/social issues~\cite{smith2013civic}. Based on our initial test, this form of outreach may yield more civic participation. However, Botivist's Solidarity strategy obtained participation from only 8\% of the volunteers  targeted. With inadequate strategies, people's civic participation is likely to be inhibited. 

Figure~\ref{fig:volunteers}b presents the number of replies per Botivist tweet (reply rate). Reply rate is a common measure in the study of participation by online audiences, and has also recently been used to study how online audiences interact with advocacy organizations~\cite{Savage:2015:PMA:2675133.2675295}.  For Botivist, a perfect reply rate of 100\% would be one reply per tweet sent, even if the tweet contained multiple user mentions, as in the call to action. We found that there was a significant difference between number of replies across strategies (a ANOVA test gave [$F(3,932)=8.594$, $p-value<0.001$]). 
The direct strategy had the highest reply rate (81\%), receiving almost one reply per tweet sent.  

Although the direct strategy reply rate appears high, it is lower than that of crowdsourcing companies such as Innocentive,\footnote{\url{https://www.innocentive.com/}} which lets people pose problems for anyone to solve and gives cash awards for the best proposals. Innocentive receives approximately 20 replies (proposals) per posed problem~\cite{WikipediaIno}. However, a stark difference between this platform and Botivist is their monetary compensation. While Botivist relies purely on volunteers, prizes for the best proposals on Innocentive can range from \$5,000 to \$1 million dollars. We think there is value in exploring non-monetary ways to generate discussion and proposals from the crowd, especially for activist groups who may have limited resources.

The reply rate of Botivist's other strategies was not as high compared to the direct one. The gain and loss strategies produced reply rates similar to product reviews~\cite{nichols2013analyzing}, while the reply rate for the solidarity strategy was even lower (21\%).

Figure \ref{fig:volunteers}c shows for each strategy the number of retweets and favorites Botivist received per tweet it sent (interaction rate). There was a significant difference between number of retweets and favorites across strategies, show by an ANOVA test [$F(3,932)=3.045 $, $p<0.01$].

Tweets following the solidarity strategy, particularly the solidarity quote or phrase shared, received the most retweets and favorites from volunteers. The rest of the strategies received almost the same amount of interactions. However, despite its high interaction rate, the Solidarity strategy received the lowest number of replies from volunteers. Interestingly, this strategy appears to prompt volunteers to retweet and favorite content; but not necessarily to make contributions in the form of replies. 

Figure \ref{fig:volunteers}d shows the interaction rate per volunteer tweet, i.e., how much volunteers retweet or favorite each others' content. There was a significant difference between volunteer interaction rate between strategies (a ANOVA test gave [$F(3,492)=19.34$, $p<0.001$]).
Here the Direct and Loss strategies  were the ones who harvested the most interactions among volunteers. We also see that people retweeted or favorited volunteers' content more than content original to Botivist. This result recalls observations of the Scratch online community in which users valued interactions with humans more than with computers~\cite{Monroy-Hernandez:2011:CCG:1978942.1979452}.

\subsection{Method: Analysis of Volunteer Responses}
Our goal was to understand the type of volunteer responses (including retweets, favorites, and mentions in addition to replies) that systems like Botivist can trigger. We were particularly interested to see if Botivist would prompt people to discuss and provide proposals for social problems. We used Upwork\footnote{Upwork is an online platform to contract Freelancers to perform tasks online. Upwork is available at \url{www.upwork.com/}} to hire three  Spanish-speaking, college educated individuals to independently classify Botivist's volunteers into people who contributed proposals/discussions relevant to a social problem, and people who contributed to off-topic conversations. We instructed these response coders to classify volunteers instead of single tweets because independent tweets can be harder to interpret.  First, two coders read through the 423 tweets of the 175 volunteers, and classified each volunteer into either on-topic volunteer or off-topic volunteer based on whether the person contributed proposals or discussion on the assigned social problem. 
The two coders agreed on the classification of 141 volunteers (Cohen’s kappa = 62: Substantial agreement). We subsequently asked a third coder to act as a tiebreaker in cases of disagreement.

\subsection{Results}
We found that the vast majority of volunteers (81\%) responded to Botivist's call to action with discussions and proposals relevant to the social problem they were assigned. A sample of these proposals includes:\emph{``Corruption isn't fought with street rallies! It's fought by being tough on crime, having honesty \& transparency!''}

Only 19\% of all of Botivist's volunteers contributed solely to off-topic discussions. The word most used in these off-topic discussions was: \emph{``bots.''}
A sample off-topic tweet:\emph{``Sorry ... I can't collaborate with bots. I have a cultural bias.''}

\begin{table}[htdp]
\begin{tabular}{l*{6}{c}r}
Strategy & Percentage Of On-Topic Volunteers\\
\hline
Botivist (all strategies) & 81\%\\
Loss & 74\%\\
Gain             & 89\%\\
Solidarity           & 82\%\\
Direct & 94\%\\
\hline
\end{tabular}
\caption{Overview per strategy of its percentage of on-topic volunteers. The Direct strategy had the most on-Topic volunteers.}
\label{table:offTopic}
\end{table}
To understand the types of actions promoted by each strategy, we show the percentage of volunteers who responded to the call to action explicitly with what was requested. Table \ref{table:offTopic} presents the percentage of volunteers who contributed proposals or discussions relevant to their assigned social problem, split out by the strategy Botivist used toward them. 

The Loss strategy had the largest percentage of volunteers who responded to the call to action with off-topic replies. Over 60\% of these replies used the term ``bot.'' Upon manual inspection it appeared that the replies were mainly questioning whether autonomous agents should help solve social problems. 
 The Direct strategy, on the other hand, had almost all of its volunteers dedicated to discussing/brainstorming solutions to their assigned social problem. 
 
In laboratory settings of face-to-face discussions~\cite{tversky1986rational}, the Loss strategy was more effective than the Direct strategy.
However, our results indicate that online behavior follows the opposite pattern. This might help shed light on how messages change when adopted by technology.  In the case of Twitter, messages used in the Loss strategy prompted people to question the participation of bots in activism. Bots with an obvious agenda might be perceived as suspect and unwanted, whereas people with an agenda would be more easily tolerated and even respected. 

\begin{table*}[htdp]
  \centering \small
  \begin{tabular}{p{7.0cm} p{5.0cm} p{5.0cm}}
    \hline
   {\bf Category}&{\bf Sample Hashtag in Tweet}&{\bf Sample Twitter \& Profile Description}\\\hline
{\bf Activism:} covers hashtags fighting for a social cause, or twitter handles of people who appear to participate in activism. & \emph{``{\bf \#WeAreAllAyotzinapa} We continue to fight against impunity. We won't stop, let's take this to the streets!''}
&\emph{{\bf@antireforma}: ``REVO... REVOLUTION.! DAUGHTER OF THIS EARTH! LOVER OF LIFE! I FIGHT FOR TRUTH, JUSTICE AND VENEZUELA!!!''}
\\\hline
{\bf Politics:} covers twitter handles of people who appear to officially be part of the political system or hashtags about political figures& \emph{``Bronco \& Fernando Elizondo (Mexican political candidates) will make history in Nuevo Leon {\bf \#TweetForElBronco}''}
&\emph{{\bf@epn}:``President of the United Mexican States.''}\\\hline
{\bf News:} covers hashtags used to give news alerts, or twitter handles of people who appear to be news reporters.& \emph{``Car crash in Central Street, 2 people hurt. {\bf\#Verfollow}''}
&{\bf @epigmenioibarra}:\emph{``I am a news reporter and producer.''}\\\hline
{\bf Marketing Analytics:} covers hashtags used to analyze one's Twitter followers to better market oneself or Twitter handles of people that give advice on how to better market oneself.& \emph{``Over 5\% of my followes have retweeted me in the last 24 hrs and 20\% in the last month {\bf \#usefulTweet}''}
&{\bf @umspromotions}:\emph{``Nigeria's top-class Promo-Marketing media outfit: Digital/Social Media Mgmt, Radio/TV Plugin, DJ Mixes, Strategic PR.''}
\\\hline
  \end{tabular}
  \caption{Description and examples of the categories that describe the groups' profiles.}
  \label{table:acti}
\end{table*}
\subsection{Method: Communication Patterns of Volunteers}
Our objective here was to understand whether those who respond to Botivist communicate on social media differently than those who decide to not reply. It might be that people who discuss certain social issues are more prone to interact with activist bots. To this end, we examined targeted users' other tweets. 
We divided the people Botivist called to action into two groups: responders and non-responders. We analyzed the public tweets from people in each group to identify any differences in content.


{\bf Text-Based Differentiation.} To obtain a descriptive assessment of what characterizes these groups beyind their response to Botivist, we use as a corpus the 200 tweets of each targeted user right before Botivist called them to action \footnote{We used data prior to Botivist's interaction, since our system might have influenced people's future conversations~\cite{aiello2012people}.} Within this corpus, we first identified the words that differentiate each group, and then sorted these words into different semantic categories.

To identify each group's key terms, we used a Mann-Whitney rank test~\cite{kilgarriff2001comparing}. This measure highlights the words which have been used more by one group than the other, per document. For our purposes, a document is all the tweets of an individual group member, and the corpus is the collection of the entire group's tweets (responders vs. non-responder). For each corpus, we calculate each word's Mann-Whitney $\rho$ measure, indicating how frequently the group has used that word, and also giving more weight to words which one group uses and the other group does not. This measure allows us to have corpora of different sizes. We then ranked the words of each corpus based on their Mann-Whitney $\rho$ score, and identify the top 1\%. These are the group's ``key terms.'' In the current work, we found that most key terms were hashtags and mentions to other Twitter users.

Next, we characterized each group's key terms using qualitative content coding. For each group, one of our authors first read a training set of 200 randomly selected tweets mentioning one or more of the key terms. This helped the researcher understand the context in which these terms were used. This author also looked up the Twitter profile of the users mentioned, since users' profile descriptions can also provide context for the interaction. This author then began to extract categories describing the distinguished words of both groups. Another author then analyzed the emerging categories and helped to adjust them. Finally, we looked at a set of 400 randomly selected tweets with key terms from both groups and produced a final list of mutually exclusive categories. Three Spanish-speaking, college educated colleagues who had not been exposed to this work were hired to categorize the key terms of each group. For each term, we showed the tweets in which it appeared, and in case the word was a Twitter handle, we showed also the user's profile description. We asked these two coders to categorize each of the 155 terms using the categories listed in Table~\ref{table:acti}. Coders selected the ``most relevant'' category for each term given its related information. The two coders agreed on 143 terms (Cohen’s kappa = .79). For the remaining 12 terms, a third coder was asked to act as a tiebreaker.

\subsection{Results}
We found that across strategies the people who responded to Botivist tended to consistently use more hashtags related to social causes, and referenced activists in their tweets. Over 50\% of the key terms of this group were hashtags and user mentions related to activism.  For instance, the hashtag \emph{\#porEsoPropongo}\footnote{\url{http://www.poresopropongo.mx/ }} (\emph{\#that'sWhyIPropose} in English) was one of the most tweeted by the people who replied to Botivist. This hashtag has been adopted by citizens in Latin America to make proposals for fighting corruption and impunity globally and locally. In addition to activism, these individuals also frequently referred to politicians and news reporters (25\% of their key terms belonged to news and 19\% to politics). However, they made fewer of this kind of reference than non-responders to Botivist. The latter tweeted more consistently about high profile politicians and news media (43\% of all their key terms were about politics and 40\% about news.) For instance,  they frequently mentioned possible Argentine presidential candidate Jose M. de la Sota (@delasotaok) or Mexico's president (@epn). They rarely used hashtags about social causes (only 6\% of their key terms).

A somewhat unexpected result was the presence of hashtags and users related to marketing analytics in the pool of responders to Botivist. For instance, these users frequently mentioned the account @m\_g\_w\_v, which describes itself as a guide to help others market themselves to obtain more followers and retweets. Since Botivist's purpose is to test message framing, it may be more natural for those who are interested in measuring and analyzing responses to content to respond to persuasive online bots.

Note however, that marketing analytics hashtags accounted for only 4\% of this group's distinct words (compared to less than 1\% for those who did not reply). Overall, the people who replied to Botivist mainly used hashtags related to activism. Table \ref{table:manWhi} presents the top 3 key terms for each group and strategy, according to the Mann-Whitney $\rho$ test (most distinct words). 

\begin{table}[htdp]
\small{
\begin{tabular}{p{0.7cm} c c c}
{\bf Strategy} & {\bf Responders Key Terms}& {\bf Non-Responders Key Terms}\\
\hline
Loss & \small @antireforma~(activism)&@epn~(politics)\\
& \small \#poresopropongo~(activism) &@delasotaok~(politics)\\
& \small @umspromotions~(marketing) & @sanchezcastejon~(politics)\\
\hline
Gain & \#cambioclimatico~(activism) & @albert\_rivera~(politics)\\
& \small @jaimerdznl~(politics)&@alvarouribevel~(politics)\\
& \small @latati2~(activism)&@eldiarioes~(news)\\
\hline
Solidarity & \#blacklivesmatter~(activism) & @juanorlandoh~(politics)\\
& \small \#ficrea~(activism)& @presidenciamx~(politics)\\
& \small @maracayactiva~(news) &\#radarparlamentario~(news)\\
\hline
Direct & \#handicapped~(activism) &@ccifuentes~(politics)\\
& \small @aristeguionline~(news) &@periodicovzlano~(news)\\
& \small \#mgwv~(marketing)&@sumariumcom~(news)\\
\hline
\end{tabular}
}
\caption{ Top 3 key terms found consistently in the tweets of people who replied to Botivst versus those who did not reply.} 
\label{table:manWhi}
\end{table}

\section{Discussion}
Our experiments demonstrate the potential of using online bots to call volunteers to action on social issues.
The majority of people called to action by Botivist made relevant contributions to the discussion,  and even started to interact with each other to further drive collaboration. Our study provides insight into the deployment of platforms that use online bots to call volunteers to action for activism, as well as demonstrating their  feasibility for social action groups that wish to expand their social media outreach. In this section, we will discuss the benefits and limitations of the Botivist system, and design implications for future systems of automated activist communication.

{\bf Feasibility}

Botivist's direct strategy yielded a reply rate of over 80\%. 
This strategy also produced a higher percentage of relevant responses (94\%). 
This high response rate and promising collection of relevant information showcases the feasibility of using online bots to help activists engage with new potential members. Our results also highlight how online bots enable community scaffolding around social issues. 

The differences between strategy groups in our data show that message framing makes a difference in terms of how much people participate in a Twitter discussion. This matches HCI findings that the way a computer presents its voice influences people's responses~\cite{nass1994computers}. However, we were surprised to discover that strategies that work well face-to-face were less effective when used by Botivist. This goes against the general trend for users to prefer computers that act ``more human''~\cite{nass1994computers}. 

Integrating solidarity quotes has long been a useful technique that marketing experts use to increment preference for items~\cite{shen2010mitigating}, and which activists have also adopted to prompt people to action~\cite{fleishman1980collective}. This strategy has also been effective online. For instance, some citizen journalists in the drug war added a solidarity component to their news reports to ensure participation to their calls to action~\cite{Savage:2015:PMA:2675133.2675295}. Political activists in Palestine also share solidarity messages on Facebook to mobilize support~\cite{wulf2013fighting}. But when adopted by Botivist, the technique resulted in the lowest number of responses. Similarly, the loss strategy is known to trigger people to produce more quality work~\cite{tversky1986rational}. In the 2011 Egyptian uprising, activists on social media used the technique, stating how a lack of action would result in more corruption, which drove more effective participation~\cite{hamdy2012framing}. But when used by Botivist, the strategy resulted in the lowest number of relevant replies, and seemed to make people actually question the role of bots in social issues. 

For bots to be effective, it is necessary to understand the communities in which they are deployed. We identified the people who responded to Botivist are individuals involved in online activism and marketing. They mentioned hashtags and Twitter accounts related to social causes and marketing analytics. Since Botivist made no attempt to disguise itself as a human user, it is likely that people linked Botivist to online marketing schemes. Therefore, the individuals who responded to Botivist were those who already found it natural to engage with marketing agents. However, the persuasiveness of the strategy also influenced people's contributions. People responded more to activist bots that did not appear to  have a ``secret agenda'' and are transparent about their purpose. 

The proliferation of persuasive technology in social media~\cite{forelle2015political} may effect how people react to certain strategies, perhaps driving a preference for less ``manipulative'' tactics. 
On the other hand, the fact that simply changing a bot's message changed subsequent response rates, without needing any interface modification or design of complex bot behavior, suggests that designing adequate dialog for online automated agents can have a strong effect on increasing volunteer participation in activism.


{\bf Generalizability}

Botivist can be adapted to focus on any social issue for which a message can be formulated. Activists have the opportunity to use Botivist to crowd-source solutions that improve education, health care, and economic conditions. City planners could use systems like Botivist  to call to action experts, stakeholders, or minorities to design novel, more inclusive urban spaces. 

{\bf Bootstrapping}

Previous volunteer systems have identified the difficulties in finding participants~\cite{brady2015gauging}. Botivist shows the power of bootstrapping onto existing networks of people to call volunteers to action using automated agents. However, if such automated social actors were to become popular, a large number of these bots might compete for people's time and attention. It would be useful for social media platform developers to enable functionality for these bots via APIs. For instance, bots could be notified of which users have already been targeted by other agents, to eliminate duplicating volunteering requests, and reduce users' ``compassion fatigue.'' Additionally, if a strategy is known to be ineffective, the system could recommend a different strategy to the bot. 

{\bf Maintainability}

Our results show that a reasonable proportion of people were willing to make relevant contributions to Botivist's causes for free. This is meaningful because cost is a key problem in maintaining systems that rely on crowds.

However, the long-term success of systems like Botivist may require automated social agents that can continuously engage the volunteers they have recruited. Such systems should take into account that while they do need to maintain contact with volunteers, they should not ping them so frequently that volunteers tune them out. 

Additionally, by mentioning different individuals in a tweet, Botivist enabled strangers to meet and start collaborating. Future work could also analyze how to design for long-term cooperation and even maintaining lasting relationships among volunteers. In particular, researchers could study designs that follow up with volunteers about their experiences, recognize them for their work, help them
feel part of the activist group, and build a sense of community~\cite{voida2012bridging}. 

On the other hand, some activists follow ``syntax rules'' when writing tweets to ease coordination~\cite{Starbird:2011:VSD:1978942.1979102},  Botivist could also  motivate standardized contributions that use a certain format or hashtag. This might also help long-term collaborations. 
%



\section{Conclusions and Future Work}
This paper introduced Botivist, a system that uses Twitter bots to engage volunteers in a discussion about social issues. The system uses the bots to quickly probe different strategies for calling volunteers to action. People on Twitter responded to Botivist's calls to action and provided relevant contributions to the discussion about corruption and impunity that we used as Botivist's test case. Volunteers also began to interact with each other, showing the potential for using bots to scaffold collective efforts. We found that the strategy which Botivist uses to call volunteers to action does matter. Tactics known to be  effective when used by humans were not as effective when adopted by online bots. A direct strategy was the most effective for getting relevant participation. We believe that the growing number of persuasive online  agents~\cite{forelle2015political,mexElec} has made people prefer interactions with greater transparency and a less ``manipualtive'' approach. Nevertheless, our experiments highlight the real-world feasibility of using online bots to call volunteers to action.

The insights from this work are limited by the methodology used and population studied. For example, those targeted by Botivist were all Spanish speakers, likely with their own cultural biases towards automated accounts (especially given the recent accusations that Latin American governments have used online bots to change political discussions~\cite{persaMersa}). Our results might therefore not generalize to an English-, Russian-, or Chinese-speaking population. Future work could study how people across the globe engage with bots to better understand the phenomenon. Our methods also focused on breadth instead of depth. Future work could conduct longitudinal studies and engage in in-depth interviews with volunteers. 

{\bf Acknowledgements.}  Special thanks to Mako Hill, Dario Taraborelli  and William Das for their great feedback in this work.

\balance

\small
\bibliographystyle{acm-sigchi}
\bibliography{sample}
\end{document}